# A BASIC GESTURE AND MOTION FORMAT FOR VIRTUAL REALITY MULTISENSORY APPLICATIONS

Annie Luciani, Matthieu Evrard, Damien Couroussé,
Nicolas Castagné, Claude Cadoz, Jean-Loup Florens

*ACROE & ICA laboratory,*
*INPG, 46 av. Felix Viallet, Grenoble, France*
*annie.luciani@imag.fr, matthieu.evrard@imag.fr, damien.courousse@imag.fr,*
*nicolas.castagne@imag.fr, claude.cadoz@imag.fr, jean-loup.florens@imag.fr*

Abstract: The question of encoding movements such as those produced by human gestures may become central in the coming years, given the growing importance of movement data exchanges between heterogeneous systems and applications (musical applications, 3D motion control, virtual reality interaction, etc.). For the past 20 years, various formats have been proposed for encoding movement, especially gestures. Though, these formats, at different degrees, were designed in the context of quite specific applications (character animation, motion capture, musical gesture, biomechanical concerns…). The article introduce a new file format, called GMS (for 'Gesture and Motion Signal'), with the aim of being more low-level and generic, by defining the minimal features a format carrying movement/gesture information needs, rather than by gathering all the information generally given by the existing formats. The article argues that, given its growing presence in virtual reality situations, the "gesture signal" itself must be encoded, and that a specific format is needed. The proposed format features the inner properties of such signals: dimensionality, structural features, types of variables, and spatial and temporal properties. The article first reviews the various situations with multisensory virtual objects in which gesture controls intervene. The proposed format is then deduced, as a mean to encode such versatile and variable "gestural and animated scene".

## 1 INTRODUCTION

Gesture, motricity, and haptic perception are deciding factors in the way we act on and we perceive our environment. During the last few years, many research centers have focused on gesture control and movement analysis or synthesis, in domains as various as surgery, aeronautics, multimedia, artistic creation, and generally speaking in every interactive systems. More and more, applications are developed in opened and versatile contexts and 3D images synthesis are produced in interactive multimedia installations, in a such a way that they have to be correlated to sound synthesis processes, or other processes.

Thus, one today would need a low level format to be used by developers and users in Computer graphics, Computer Music, interactive systems and versatile VR platforms. Several formats already exist for encoding motion control/gesture data. However, these formats were introduced in the context of quite specific application and uses, usually do not focus strictly on motion and gesture, and were not designed, at first, to be used in versatile and heterogeneous contexts. The data they encode are thus context dependent, and they hardly correspond with motion data in its more general meaning. This article questions the need of a minimal and low level description of gesture and motion data, able to code all their features independently of the context in which they have been produced and will be used.

The two next sections situate the various notions at hand and discuss the properties of motion/gesture data. The section 4 reviews the various existing file formats, and discuss their advantages and lacks. The two last sections present the requirements for a more generic encoding of gesture data and the main features of the proposed format, called GMS.

## 2 AMBIGUOUS TERMS

Given the ambiguity of various terms the article focuses on, a few definitions have to be discussed.

### 2.1 Action / Movement / Signal

*Movement* (or *motion*) is the moving in space of a part or of the totality of a system. Movement should be distinguished from *action*, which is the result of the movement. Following (Smith & al., 1984) , action is the result of the task achieved or to be achieved - for example: "to drink a glass of water". Each action can be performed by means of many different movements. And a movement can be characterized by properties, such as soft, vivid, etc.

Action can be described at a high symbolic level, for example by language, or by means of event-based representations. Conversely, movements are explicit spatio-temporal phenomena. They need to be represented as temporal signals.

### 2.2 Motion and Gesture

Another evidence concerning motion is that it refers to the evolution produced by a physical system, whatever it is: human body, real mechanical objects equipped with sensors, virtual objects, etc. One can speak as well of the motion of a human body, of a leaf, of a sounding source, etc. Motion is therefore considered as the result of the performance, i.e. as an output of an evolving system.

Conversely, when focusing the use of such movements as "a command" or an input to control the motion of another system, the only usable terms are "action" or "gesture". Unfortunately, as said in the previous paragraph, "action" refers more to the high level of the task than to way it was achieved. It is very general, and can be used for any type of systems. Conversely, "gesture" is usually reserved to humans. Facing this dilemma, we decide to emphasize the differentiation between input and output functionalities, more than the nature of the evolving system that produces the signals. Thus, we call "gestures" all the motions that can be applied as an input signal, i.e. as a "cause of a performance", whatever the producing system is (human beings or other objects).

Anyhow, Motion/output and Gesture/input are two representatives of similar temporal signals, corresponding with an evolving physical system – see Figure 1. Thus, to simplify, the term 'gesture' and 'motion data' could be used in the following to refer to this sort of 'gesture like' data, no matter it is considered as an input or an output of an evolving systems, possibly a human body.

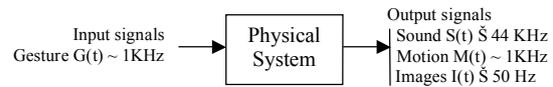

Figure 1: Gestures and Motion as input and outputs of physical systems. Gesture (input) and Motion (output) are comparable types of data.

Rather than discussing the differences between Gesture and Motion, one can emphasize their similarities. This leads to identify them as a new and unique *type of data*, beside visual and acoustic data, and to introduce a dedicated encoding format.

### 2.3 Gesture and Control

Gestures can be considered as a common feature of numerous situations, applications, systems etc., and they may be a privileged mean for allowing those systems to cooperate: animation control, animation performance, musical control and performance, multisensory control and performance such as in VR or in multimedia arts.

The figure 2 illustrates some basic cases of gesture signals and gesture control. On the figure, grey circles are placed on the streams one can consider as gesture streams (or as being equivalent to a gesture signal). Conversely, the grey squares are placed on the streams that are no more gesture streams.

In the case of gesture control of real or digital musical instruments, the gesture signals can be either the signals sensed directly on the human body (fingers, hands, arms, whole body) (figure 2, grey circles 1 and 2), or the motions produced by an object manipulated by the player (hammer, key, bow, stick, etc...) (figure 2, grey circle 3).

In the case of gesture control of 2D or 3D visual motions, the gesture signals can be either the signal sensed directly on the human body (figure 2, grey circles 4 and 5), or the motions produced by an object manipulated by the performer (sticks, force feedback devices, etc.) (figure 2, grey circles 6 and 7). Conversely to what happens in musical instruments, in which the output "sound" is usually not considered as a "gesture", the 3D motions provided by a 3D virtual objects (Figure 2, grey circle 7) are of the same nature as the 3D motions produced by a human body in body motion capture. They may be used as an input for controlling another system (3D object , musical instrument, etc.). Consequently, a 3D virtual object produces signals that can be considered as being homogeneous with "gestures" (Figure 2, grey circles 4, 5, 6).

In the case of the gestures that produce formal and symbolic data, as when acting on a computer keyboard, the gesture signals (Figure 2, grey circles 8 and 9) are transformed in outputs (words, sentences) that can usually no more be considered as gesture signals.

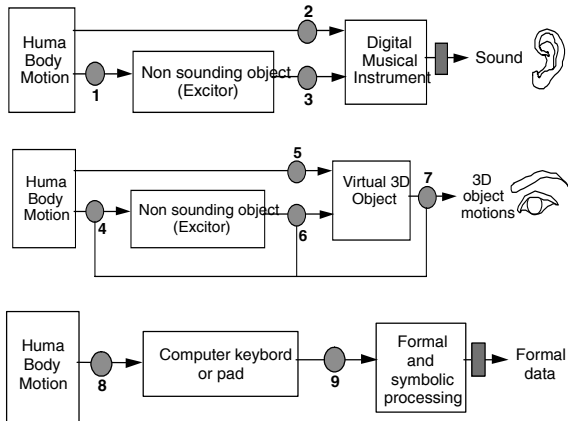

Figure 2: Various cases of gesture control. Up: gesture control of digital musical instrument. Middle: gesture control of a 3D virtual object. Down: gesture control of formal and symbolic data.

## 3 CHARACTERIZATION OF GESTURE SIGNALS

Gesture signals, whatever the way they are produced (objects or human motion, virtual object…), and whatever the way they are considered (as outputs or inputs of evolving systems), do present specific properties that allow distinguishing them among other temporal signals (especially aero-acoustical signals or visual signals). This section reviews these properties that will be used as a basis for defining a basic generic format.

### 3.1 Morphological Versatility

A first evidence is the morphological versatility of gestures. While images and sounds can be displayed in predefined environments (displays or 3D Caves, with fixed sizes and resolution, stereo or quadraphonic rendering for sounds, etc.), the structure and the morphology of gesture signals are more versatile, depending on the tasks and the manipulated tools. To take into account this inner versatility, we propose to structure gesture signals along two complementary features: *geometrical* and *structural dimensionalities*.

### 3.1.1 Geometrical Dimensionality

Geometrical dimensionality refers to the dimensionality of the space in which the gesture is evolving. It can vary from a pure scalar or a set of scalars to geometrical 1D, 2D, 3D.

For example, piano or clarinet keys are pushed or closed according to a 1D finger motion (figure 1a). More generally, controlling sounding objects or tuning parameters (for example the value of an elasticity or the amplitude of a deformation), can be made through devices that evolve in a 1D non oriented space, like set of sliders or set of knobs (figures 1c, 1d), and that can be described by a scalar or a set of scalars.

Conversely, in cartoon animation or in scrap-paper animation or animated painting, the space is reduced to a plane. Gestures and motions evolve in a 2D space (Figure 3, g), and can be described on two orthogonal oriented axes.

When we manipulate an object (real or virtual), the dimensionality of the space is obviously 3D. Thus, describing motion or gesture requires using three orthogonal oriented axes (figure 3, e, f, h).

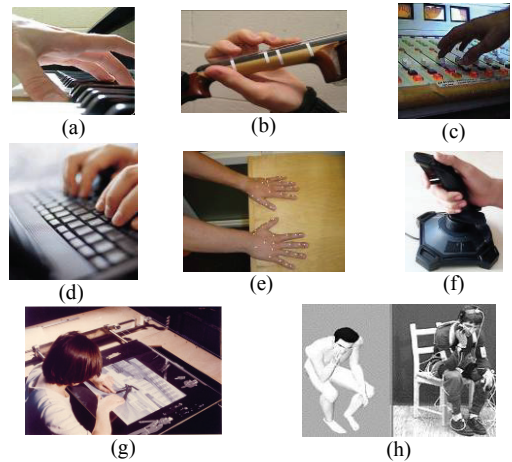

Figure 3: Versatility of the gesture morphology.

### 3.1.2 Structural Dimensionality

For a given geometrical dimensionality, gestures have another type of variability, related to their functional organization. We call this second variability *structural dimensionality*. In the case of articulated 3D solids, it usually corresponds to the number of degrees of freedom. A couple of examples allow illustrating the variability of structural dimensionality of gesture.

When we are acting on a keyboard of n keys (a piano keyboard, a computer keyboard, a set of buttons…), the performed gesture, and similarly the signals produced by the n keys, the geometrical dimensionality is 1, but the system can be considered in two ways either as n independent systems (for example n systems of one key), or as a single system of n degrees of freedom.

When dealing with human body motion, the geometrical dimensionality is 3 but the structural dimensionality (ie: the number of considered degrees of freedom) is variable: more than 200 in the real body, only N when using a motion capture systems with N sensors (N=1 to 16, or more).

In physical models of a bowed string, the two dimensions of the deformations (pressing the string, bowing the string) are usually decoupled, and the system can be considered as two superposed 1D gestures, thus as one system of two 1D DoF.

In particle physical model of a flow, the number N of 3D particles, i.e. the number N of 3D DoF, could be more than some hundreds. Considering that this flow can be used to control large spatial scene (for example wind playing a musical instrument, flow controlling atmospheric or oceanic deformations), the motions can be stored as gestures of N 3D DoF.

### 3.1.3 To Conclude

This analysis and these examples show that the geometrical dimensionality of gestures should not be fixed, since it can vary from scalar to 3D (we don't examine here space with larger geometrical dimensionality, although the case can occur). They show also that the structural dimensionality can vary widely (from 1Dof to several hundred). In addition, the structural dimensionality cannot be totally independent of how the system is analyzed by the designer that must be able to declare them freely.

## 3.2 Quantitative Ranges

Additionally to the geometrical and structural dimensionalities, gesture signals present specific quantitative spatial and temporal features.

### 3.2.1 Temporal Ranges

As shown on Figure 4, the frequency bandwidth range is one of the quantitative features that allow distinguishing gesture signals among others (especially aero-mechanical signals, visual motions).

Although the three zones of the figure 4 are overlapped, they enable a useful categorization: visualizing motions requires a sampling rate up to 100 Hz; manipulating an object with force feedback requires a sampling rate from some Hz to some kHz; recording sounds requires a sampling rate from around 10 kHz to some 40 kHz. Indeed, gesture signals are at the middle range.

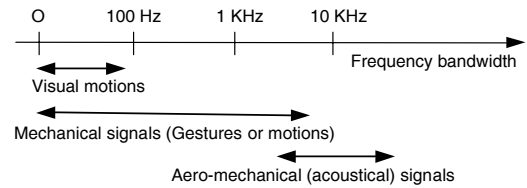

Figure 4: Temporal range of sensory signals.

### 3.2.2 Spatial Ranges

Spatial positions and ranges is another feature of gesture signals. Whereas audio signals are zero-centered deformations of about some millimeters, gestures and visual motions correspond with deformations and displacements that are non-necessarily centered on 0, and that are at a larger spatial scale, from centimeters to meters.

### 3.2.3 To Conclude

One can note that the quantitative properties of gesture signals position them at middle range of the signals involved in interactive computer graphics, specifically when using force feedback devices and physically-based simulation, or multisensory VR simulation platforms including sounds: (1) it is similar to visual motion as for spatiality, but it needs higher frequency rate; (2) it needs a lower frequency rate than the sound, but it runs at higher non-centered spatial range.

Indeed, one can consider the gesture signal as a pivot for multisensory VR simulation platforms. Visual motions can be produced by simply under-sampling them. A non-linear control-simulation process can then produce sounds, transforming low frequency gesture command signals in high frequency sound signals.

## 3.3 Type of variables

Since motions and gestures are produced by physical systems (either real or virtual), and used to control physical systems, the data can be of two different types: (1) extensive variables, such as variables that derivates from spatial information (positions, velocities, angles…); and (2) intensive variables as forces, torques, etc. Noticeably, one can note that visual and acoustical data are only extensive variables (positions and/or displacements).

In natural situations, when gestures are used for manipulating object, physical energy is exchanged between the two interacting bodies (for example object and human). To represent such interactive dynamic systems, the correlation between extensive and intensive variables must be considered, either explicitly as in Newtonian formalism, or implicitly as in energy formalisms. Both extensive and intensive variable are needed for a complete encoding of the interaction.

# 4 A REVIEW OF EXISTING MOTION FILE FORMATS

Several motion file formats have been developed so far (Meredith, 2000 , Van Baerle, 2000). Here, we position some of the most known (see Table 1) along the properties of gesture signals elicited in the previous paragraphs: versatility of the geometrical and structural dimensionality, spatial and temporal requirements, and multiplicity of the types of data.

Table 1: some of the most known 'gesture/motion' format.

| BVA, BVH ASK, SDL | [BVH] Developed by Biovision, a firm providing optical motion capture systems. |
|---|---|
| HTR, TRC | [HTR] Developed by the Motion Analysis. |
| AOA | [AOA] Designed by Adaptive Optics, a firm that develops motion capture hardware. |
| ASF AMC | [ASF] Developed by Acclaim, a video game firm. Now used by Oxford Metrics (Vicon motion capture Systems). Public domain. |
| BRD | [BRD] Introduced by Lambsoft, for the "Flock of Birds" motion capture system. |
| CSM | [CSM] 3D Studio Max and Character Studio format. Recoding of characters' movements. |
| C3D | [C3D] National Institute of Health. Initially encoding movements of athletes. Largely used today (ex: as a file format for encoding basic marker data, for the Vicon Motion Systems). |

## 4.1 Origin of the Existing Formats

The considered formats have been historically introduced in the context of specific application and uses – mostly in the context of Motion Capture. Though some have eventually been adapted to other applications and usages than their preliminary target, they had neither been designed at first to be used in other applications, nor as a data exchange mean. Consequently, the data they encode, and the manner these are encoded, depend on the context of the usages preliminary foreseen, and hardly correspond with gesture / motion data in its more general meaning.

## 4.2 Object Dependence

Except C3D, all the formats imply a hierarchy depicting a hierarchical dependency over the various recorded signals. The hierarchy is inherited from the hierarchical structure of the object that produces the signals. For example, describing the movement of a humanoid is made according to a pre-defined tree-like structure of a cinematic chain of the body (Herda &al., 2000). In that case, most often, the value of a position/rotation is given in the coordinate system of the upper hierarchical level it belongs to.

As for it, the CSM file format developed along with the Character Studio software is not explicitly hierarchy-based. It encodes the motion of a set of 3D points in absolute coordinate. However, using CSM files within Character Studio requires the names of all the points to match Character Studio setup. A skeleton is implied, and the number of points and their 'meaning' are fixed.

Such a hierarchical structure has proved to be operational in domains such as motion planning or character animation. It is most often adapted to the case of articulated objects such as human body, rigid-articulated robot, etc. However, the use of a hierarchical structure does not apply to motion/ gesture encoding in its more general meaning – especially to the case of deformable non tree-articulated objects and mechanisms.

## 4.3 Context Dependence

Some of the discussed formats contain additional data dealing with the context of data acquisition. The context data are typically data concerning the parameters of the capture system (number of sensors, captors used to determine the movement at a certain moment, sensitivity of sensors, etc.).

The C3D file format is the most representative example of such additional data recording. In a C3D file, each motion stream is associated with the description of the measurement process that was used. Indeed, C3D corresponds firstly with measurement or experimental setup recordings.

## 4.4 Predefined Dimensionality

Most of the formats predefine a specific dimension for the recorded data: 3D points in AOA or CSM, "7D points" in BRD. In the BVH, ASF/AMC and HTR file formats, the dimension of each join is declared in accordance with the constraints underlain by the hierarchical structure. Omitting a coordinate corresponds with constraints of the mechanical system. Once again, one can observe that these formats suppose that the recorded motion data are intimately associated with a tree-like structure of the object supporting the sensors. As for them, C3D files store conventional 3D positional information. 1D or 2D motion data may be recorded, by fixing one or more coordinate to zero all along the stream. Though, scalar data can also be recorded within the raw data (see §4.5).

Anyhow, one can consider that pre-defining the dimensionality of the data to be stored is a real problem since the versatility of dimensionality is a core property of gestures, and since one may often have to consider coexistence of different dimensionalities in the same gesture stream.

## 4.5 Type of Variable

C3D files can encode two categories of data: 'analog data', i.e. scalar data acquired from any kind of sensor, and '3D data' i.e. motion-like data displayed by a motion capture system. The format implies no inner limit in the type of the 'analog data', so that any type of variable may be recorded, including hearth rate, elecromyogram, etc. Additionally, a non-optimized trick allows recording 'kinematical data' (angles, moment, acceleration…) processed from the motion data. Thus, besides the fact that C3D is widely used for motion recording, it does overpass the strict scope of motion and gesture. It is a very flexible and adaptable format – but it is too general for being a Basic Law-Level format adapted to the category of data we call "gestures".

As for the type of the 'gesture/motion data' themselves, the various file formats are restricted to position-like data (C3D allow recording other types, but only within the very open 'analog data' sections). They do not natively include the possibility to encode the type of the variables. They are especially not adapted natively to the recording of intensive variable.

## 4.6 Data Coding

Except for the C3D format, ASCII encoding is used in the existing formats. This corresponds to the fact that the frequency bandwidth of motion capture systems and gesture controllers used to be on the VLF range (from 10 to 300 Hz) and that recorded gestures used to have a small number of DoF and tracks (less than one hundred). Though, today, one usually considers that the frequency bandwidth of gesture signals may be of some kHz (especially when dealing with force feedback gesture interaction). In addition, the diversification of the objects (real or virtual) producing motions, and the decrease of the costs of sensors, imply that the number of tracks to be stored is significantly increasing. Consequently, gesture files may correspond with amount of data far more important than it had been considered so far, and ASCII encoding is no more usable. Indeed, the use of a binary format (like for sound, images or movies) is a necessity for a generic gesture/motion format.

## 5 THE ™GMS FORMAT

On the basis of the previous analysis of gesture (or more generally of motions usable as a command of an evolving system), this section introduces the proposed Low Level Gesture File Format. This format is called ™GMS, which stands for "Gesture and Motion Signals" - file extension is *.gms*.

### 5.1 Requirements for a Gesture File

#### 5.1.1 Sampling Rate

First of all, a gesture signal is a signal sampled at the necessary Shannon frequency. In the current 0.1 version, a unique sampling rate is used for all the data in a GMS file. This choice does not correspond to an optimal encoding, and will be reconsidered in future versions.

#### 5.1.2 Geometry and Structure Encoding

We then propose to organize the gesture data in four levels at maximum: Tracks – Channels – Units – Scenes :

• **Gesture Track.** The track i contains a sampled monodimensional scalar $a_i(t)$ corresponding to a single A-D track $a(t)$, sampled at its Shannon rate.

• **Gesture Channels.** *Channels* support the geometrical dimensionality of gesture signal. A Channel is composed of one or several tracks. A Channel can be 1D0 (a pure scalar), 1Dx (a vector on x axis), 1Dy, 1Dz, 2Dxy, 2Dyz, 2Dzy, or 3Dxyz. A Channel can encode either "extensive variable (EV)" i.e. homogeneous to spatial variables (positions P, Angles A, velocities V, Accelerations G) or "intensives variables (IV)", i.e. homogeneous to forces F.

• **Gesture Units.** *Units* are provided for supporting the structural dimensionality of gestures.

A *unit* is then composed of several channels in which the signals are not dynamically independent, whatever the way in which they are correlated is. This differs from the structural properties of motion capture formats that focus on tree-like structures. The user can declare units freely. Along with supporting a low-level structuring of the signal, Units allow preventing non-relevant signal processing in future uses of the data, for example breaking of the correlation.

- **Gesture Scenes.** Finally, a scene is composed of several *units* that are not – or can be considered as not - dynamically linked.

### 5.1.3 An Example

In table 2, we illustrate how a user could structure a complex gesture scene like the one described in the 'DESCRIPTION' column.

Table 2: a possible structure for a gesture scene.

| UNIT | DESCRIPTION | CHANNELS |
|---|---|---|
| Pianist | One musician plays with a small keyboard composed of 8 piano-like keys. | 8 Mono dimensional Channels "PianoKeys": PK1 [EV(P), 1Dz], ... , PK 8 [EV(P), 1Dz]. |
| Stick Source | Another one is controlling the spatial position of a sound source via a 3D stick. | 1 3Dxyz Channel "SoundSource": SS [EV(P), 3Dxyz] |
| Light | Another one is controlling the orientation of a light source via a 2D force sensor (a force pad). | 1 2Dxy channel "LightSource": LS [IV(F), 2Dxy] |
| Dancer | A dancer is equipped with a motion capture system of 16 3D points. | 16 3Dxyz channels "DancerPoints": DP1[EV(P), 3Dxyz], ..., DP16[EV(P), 3Dxyz] |
| Juggler | A juggler is manipulating a 6D force feedback ball to play with a 6D virtual racket. | 1 3Dxyz channel "Ball1": BL1 [EV(P), 3Dxyz] |
| | | 1 3Drqf channel "Ball2": BL2 [EV(A), 3Drqf] |
| Fluid | A virtual fluid of N 1D masses acts on a virtual string. | n 1D "fluidMass": FL [EV(V), 1D0] |

## 5.2 ™GMS Format Implementation

This paragraph describes the current implementation of the concepts developed previously. The requirements have been implemented in the version 0.1 of the ™GMS gesture format, through a C/C++ read/write library, which is currently submitted to file formats authorities, and as a MIME type.

The format is built above the IFF (Interchange File Format) coding standard for binary files. A description of this format can be found in (Morrison, 1985). IFF, and its equivalent RIFF, are largely used by several formats as AIFF for sound files, AVI for movie files, ILBM or GIFF for picture files.

As required by the IFF coding standard, data in the proposed gesture file are encapsulated in structures called *chunks*.

Table 3 presents a description of the data contained in each of these chunks. The binary types of those data are CHAR (a byte), USHORT (a 2 bytes unsigned integer), LONG or ULONG (a 4 bytes signed or unsigned integer), FLOAT32 or FLOAT64 (a 4 or 8 bytes floating point number - IEEE 754 standard). All the data covering more than one byte in the file are encoded in BIG ENDIAN.

Table 3: description of the data contained in each chunk.

| | | |
|---|---|---|
| Header | CHAR[4] chunkId = 'FORM' | ULONG fileSize; |
| | CHAR[4] fileType = 'GSM ' | |
| Version | CHAR[4] chunkId = 'VERS' | ULONG chunkSize |
| | USHORT versNum | USHORT subVersNum |
| Scene | CHAR[4] chunkId = 'SCEN' | ULONG chunkSize |
| | USHORT sceneNameLength | CHAR*sceneName |
| | ULONG nbFrame | FLOAT64 freq | USHORT dataType |
| | FLOAT64 scale | ULONG blockSize |
| Unit | CHAR[4] chunkId = 'UNIT' | ULONG chunkSize |
| | USHORT unitNameLength | CHAR* unitName |
| Channel | CHAR[4] chunkId = 'CHAN' | ULONG chunkSize |
| | USHORT chanNameLength | CHAR* chanName |
| | USHORT dimension | USHORT type |
| Frame | CHAR[4] chunkId = 'FRAM' | ULONG chunkSize |
| | TypeOfData[nbTrack][nbFrame] frames | |

The *version* chunk corresponds to the chunk structure to be found in the file. Future evolutions of the gesture file are foreseen.

The *scene* chunk contains information about the gesture scene encoded in the file. *dataType* gives the type of data of the signal. The current version supports three type of data: FLOAT32, FLOAT64, and LONG. *scale* is a scale factor to apply on the signal. *blockSize* allows setting up an alignment of the frames (ie: the gesture data) on multiple of *blockSize*. This may be useful on some architecture to process the file in real time. If *blockSize* is 0 (default), the size of a block equals to the size of a frame.

The *unit* and *channel* chunks encode information on a gesture/motion unit and channel. *Channel* and *unit* chunks are interleaved: the current channel belongs to the last declared unit. The *dimension* integer encodes the dimension of the channel, and accordingly its number of tracks. Dimensions supported in version 0.1 are: 0D *(*pure scalar); a vector on 1Dx (resp. 1Dy or 1Dz), on 2Dxy (resp. 2Dyz or 2Dzx), on 3Dxyz. *type* encodes the type of the data. Current supported types are position, and force.

The *frame* chunk contains the signal itself, given frame by frame.

## 6   CONCLUSION

While gesture devices, especially haptic devices, develop, while applications communicate more and more though gesture-like data, the definition of a low level gesture format can be seen as a major need for the near future.

To do this, we extracted the specific properties of gesture/motion signals over other temporal signals (sounds and images): morphological versatility (decomposed in geometrical and structural properties), spatial and temporal ranges, and variety of type of data. We showed that these properties are able to explain why gestures can be considered as at "a hub place" in multisensory situations and how they can be shared by various applications. This led to introduce a basic file format, able to encode all the minimal and necessary features.

Various evolutions are foreseen, including:
- extending the format to multi-frequence gesture stream; eventually, defining a set of standard frequencies.
- extending the possible dimensions of channels, especially to 6D; although the necessity of such ND channels is not true evidence, and is still under study.
- extending the supported *types* of data, including for example displacement, velocities, angles, torques, etc.

These extensions, though, require further analysis for being exhaustive without damaging the 'minimality' and 'simplicity' of the proposed format. Suggestions and comments will be welcome.

The proposed format has been used in physically based multisensory simulators for synchronous haptic interaction with computer animation and sound synthesis, in hard and soft real-time, as well as in separated physically-based animation simulators and sound synthesis simulators. Experiences for connecting the various systems (computer animation, computer sound, multimodal systems) through the proposed gesture format are in progress in the laboratory.

## ACKNOWLEDGMENTS

This work was supported by the FP6 Network of Excellence IST-2002-002114 - Enactive Interfaces, and by the French Ministry of Culture.